\documentclass[aps,twocolumn,amssymb,prl,showpacs,superscriptaddress,10pt,floatfix]{revtex4-2}
\usepackage{comment}
\usepackage{amsmath,bm}
\usepackage{graphicx}
\usepackage{xcolor}
\usepackage{enumitem}
\begin{document}

\title{Radiation Reaction effects on Coherent Emission in Relativistic Magnetized Shocks}

\author{Yu Zhang}
\affiliation{State Key Laboratory of Ultra-intense Laser Science and Technology, Shanghai Institute of Optics and Fine Mechanics, Chinese Academy of Sciences, Shanghai 201800, People’s Republic of China
}

\author{Yuan-Pei Yang}
\affiliation{South-Western Institute for Astronomy Research, Yunnan Key Laboratory of Survey Science, Yunnan University, Kunming, Yunnan 650504, People’s Republic of China
}

\author{Liang-Liang Ji}
\email{jill@siom.ac.cn}
\affiliation{State Key Laboratory of Ultra-intense Laser Science and Technology, Shanghai Institute of Optics and Fine Mechanics, Chinese Academy of Sciences, Shanghai 201800, People’s Republic of China
}
\date{\today}

\begin{abstract}
Relativistic magnetized shocks are natural sources of coherent radiation, representing a promising framework for fast radio bursts (FRBs). This study explores how the radiation reaction (RR) effect, triggered by high-energy photon emissions during shock radiation, significantly alters particle dynamics and coherent radiation properties. Using kinetic particle simulations, we demonstrate that RR severely suppresses electron energies from shock acceleration, resulting in multiple coherent gyration cycles at the shock front. It amplifies the intensity of coherent radiation and boosts energy efficiency by several fold, as compared to the single gyration cycle in the standard model of relativistic magnetized shocks. We further find that the coherent radiation spectrum from RR-mediated shocks is characterized by upshift peak frequency, broaden bandwidth, and narrow spectral peak. These RR-induced radiation changes may be related with several observed FRB phenomena, including the statistically positive correlation between luminosity and bandwidth in repeating and one‑off FRBs, the narrow spectra seen in some FRB events, and the bimodal energy distribution reported in FRB 20121102A. 

\end{abstract}

\maketitle

Magnetized shocks are commonly present in astrophysical environments, such as supernova remnants, pulsar wind nebulae and active galactic nuclei jets. They are considered primary sites for accelerating high-energy cosmic rays and have recently been suggested as a leading candidates for powering fast radio bursts (FRBs)\cite{Lyubarsky2014, Beloborodov2017, Waxman2017, Sironi2021, Plotnikov2019, Iwamoto2024, Beloborodov2020, Vanthieghem2025_1, Bernardi2025}. FRBs are luminous, millisecond-duration extragalactic transients observed at GHz frequencies  \cite{Cordes2019, Lorimer2018, Petroff2019, Platts2019, Katz2018, Popov2018, Zhang2020}. Their extreme brightness temperature implies coherent radiation, but the emission mechanisms remain unsettled. Current theories favor magnetars as the central engines, bolstered by the Galactic FRB 200428 associated with an X-ray burst from magnetar SGR J1935+2154 \cite{Bochenek2020,CHIMEFRB2020}. The currently proposed origin models for FRBs can be broadly classified into two categories based on their radiation mechanisms: curvature radiation of electron bunches\cite{Kumar2020,Lu2020_1,Yang2018,Lu2018,Yang2020} and synchrotron maser mechanism in relativistic shocks \cite{Vanthieghem2025_1,Lyubarsky2014,Beloborodov2017,Waxman2017,Sironi2021,Plotnikov2019,Iwamoto2024,Beloborodov2020}. In shock-based FRB scenarios, relativistic magnetized shocks may be excited either by magnetar flare–wind interactions \cite{Lyubarsky2014,Beloborodov2017,Waxman2017,Sironi2021,Plotnikov2019,Iwamoto2024, Beloborodov2020} or fast magnetosonic wave breaking \cite{Vanthieghem2025_1, Bernardi2025}. Such shocks can generate coherent electromagnetic radiation through the synchrotron maser instability. Here, cold magnetized particles entering the shock are accelerated and undergo collective synchrotron-like motion at the shock front, emitting coherent electromagnetic pulses, known as precursor waves that escape upstream region \cite{Sironi2021, Plotnikov2018, Plotnikov2019, Iwamoto2017, Iwamoto2018, Iwamoto2019, Iwamoto2024, Lyubarsky2014, Vanthieghem2025_2}.


Given FRBs' enormous isotropic luminosity of $L\sim(10^{38}-10^{46})\,\rm{erg/s}$ \cite{Zhang2020,Bochenek2020,Ravi2019,Zhang2018}, the normalized electric field near the source \cite{Yang2020b} can be estimated as $a_{0}=eE_{\rm{FRB}}/m_e\omega\sim10^{4}L_{44}^{1/2}R_{10}^{-1}\omega_{10}^{-1}$ \footnote{the convention $Q_{x}=Q⁄10^{x}$ is adopted in cgs units unless otherwise specified.}, where $e$ is the electron charge, $m_e$ is the electron mass, $E_{\rm{FRB}}$ is the electric field of FRBs, $\omega$ is the FRBs' angular frequency and $R$ is FRBs' transport distance. During FRB's radiation or initial propagation, electrons are expected to be accelerated in the strong field, reaching Lorentz factor $\sim a_{0}^{2}$, and emit high energy photons. 
Such radiation could induce strong recoil on electron motion, known as the Radiation Reaction (RR) effect \cite{Zeldovich1975}. For GHz wave, the threshold for RR dominance can be estimated as $a_{\rm{thr}}\sim (3\lambda_{\rm{GHz}}/4\pi r_{e})^{1/3}\simeq2.94\times10^{4}$ \cite{Bulanov2013,Ji2014,Gonoskov2022}, where $\lambda_{\rm{GHz}}\simeq30\;\rm{cm}$ is the wavelength of 1 GHz radio wave and $r_{e}$  is the classical electron radius. In thermal plasma with strong magnetic fields, RR could induce Landau population inversion and spontaneously drive coherent radiation through cyclotron maser instability \cite{Bilbao2023,Bilbao2024_1,Bilbao2024_2}.

Here we extend the study of magnetized shock into the RR regime. Using kinetic simulations, we show that RR cooling strongly suppresses electron–positron acceleration and triggers multiple coherent gyrations in the pair flow at the shock front, in contrast to the single gyration model in standard relativistic magnetic shock theory. The modified dynamics produce distinct radiative signatures absent in RR-free scenarios: enhanced coherent emission, upshifted characteristic frequency, broadened bandwidth, and narrow spectral peak. These features are consistent with the following FRB observations: the luminosity–bandwidth correlation in repeating and one-off CHIME/FRB events \cite{Chen2022,Pleunis2021}, the narrowband spectra reported in \cite{Kumar2021,Zhang2023,Zhou2022,Law2017,Yang2023}, and the bimodal energy distribution of FRB 20121102A \cite{LiD2021}, together with its mode-dependent bandwidth variations \cite{Zhong2022}. Our study also implies that astrophysical environments naturally possess the extreme conditions required to probe the RR effect, a central topic in both classical and quantum electrodynamic (QED) realms of strong-field physics \cite{Piazza2012,Gonoskov2022,Cole2018,Mirzaie2024}. The FRB observations open up a unique and complementary avenue for investigating radiation-dominated physics.

We simulate the magnetized shock in the post-shock frame using the PIC code SMILEI-1D \cite{Derouillat2018}. As the common set-up employed in previous works \cite{Spitkovsky2008,Iwamoto2017,Iwamoto2018,Iwamoto2019,Iwamoto2024,Plotnikov2018,Plotnikov2019,Sironi2021}, the upstream is a cold, ultra-relativistic plasma jet composed of electrons and positrons, which drifts along the $-\boldsymbol{\hat{x}}$ direction with a bulk Lorentz factor $\gamma_{0}$. The shock is launched as the incoming flow reflects off a wall at $x=0$ and propagates along $+\boldsymbol{\hat{x}}$. The pre-shock jet carries a frozen magnetic field $\bm{B}_{0}= B_{0}\boldsymbol{\hat{z}}$ and an electric field $\bm{E}_{0}= -\beta_{0}B_{0}\boldsymbol{\hat{y}}$, where $\beta_{0}=(1-1/\gamma_{0}^{2})^{1/2}$. We employ 100 macro-particles per species per cell, spatial resolution $\Delta x=(1/256)\;c/\omega_{p}$, time step $\Delta t=0.99\Delta x/c$. The spatial domain ranges $(0-2000)\;c/\omega_{p}$, and the total time after the shock formation is $2000\;\omega_{p}^{-1}$. Here $\omega_{p}=(8\pi n_{0}e^{2}/\gamma_{0}m_{e})^{1/2}$ is the pair plasma frequency, $n_{0}$ is the electron/positron number density of the injected flow, respectively. Field strength of the shock is  parametrized via the magnetization $\sigma=B_{0}^{2}/8\pi\gamma_{0}n_{0}m_{e}c^{2}=\omega_{c}^{2}/\omega_{p}^{2}$ with $\omega_{c}$ being the gyro-frequency of upstream electrons.

The RR effect here can be described with the classic Landau-Lifshitz (LL) formula \cite{Landau1971},
\begin{equation}
\bm{F}_{rr}\simeq -\frac{2e^{4}}{3m_{e}^{2}c^{4}}\gamma^{2}\bm{\beta}[(\bm{E}+\bm{\beta}\times
\bm{B})^{2}-(\bm{E}\cdot\bm{\beta})^{2}]
\end{equation}
Here $\bm{E}$ and $\bm{B}$ denote the local electric and magnetic fields acting on the lepton, while $\gamma$ and $\bm{\beta}$ are the particle's instant Lorentz factor and normalized speed, respectively.

We consider a regime where coherent shock radiation operates within the GHz band while effectively inducing the RR effect. The radiation frequency in the shock upstream rest frame (laboratory frame) is given by $\nu = \gamma_0 \omega_{\text{peak}}/2\pi$, where $\omega_{\text{peak}} \sim 3\sqrt{\sigma} \omega_p$ is the semi-empirical peak angular frequency of the precursor measured in the downstream rest frame (simulation frame) \cite{Plotnikov2019,Sironi2021}. This leads to 
\begin{equation}
\nu=2.7\times10^4\sqrt{\gamma_0 \sigma n_0}
\end{equation}

\noindent{Here $\nu$ is in unit of $\rm{Hz}$, $n_0$ is in unit of $\rm{cm}^{-3}$, these units apply throughout the following text unless specified otherwise.}

On the other hand, the RR effect can be quantified by the ratio of the energy radiated by an electron to its initial energy. Over one oscillation cycle, this is governed by the parameter $R = \alpha \chi a_0$ \cite{Koga2005}, where $\alpha = 1/137$ is the fine structure constant, the QED factor $\chi$ is the dimensionless quantum parameter, given by $\chi\simeq(\gamma/E_{s})\sqrt{(\textbf{\emph{E}}+\bm{\beta} \times \textbf{\emph{B}})^{2}-(\bm{\beta}\cdot\textbf{\emph{E}})^{2}}$ \cite{Ritus1985}, with $E_{s}= 1.32\times10^{16}\,\rm{V/cm}$ being the Schwinger field strength. For significant RR in the precursor region, as shown in Fig. 1(d–f), the cumulative loss over $N$ oscillation cycles should approach to the electron initial energy. Defining $K$ as the ratio, one obtains
\begin{equation}
K=N R = N\alpha \chi a_0 \sim (0.1-1)
\end{equation}

In our simulations, electrons/positrons exhibit a significant energy reduction after propagating approximately $10\;c/\omega_p$ ($N_p \sim 10$ plasma oscillation cycles) and gradually saturate, as shown in Fig. 1(d–f). Considering $\omega_{\rm{peak}} \sim 3\sqrt{\sigma}\;\omega_p$ \cite{Plotnikov2019}, one obtains $N \sim (\omega_{\rm{peak}}/\omega_p) N_p \simeq 10^2$ over typical magnetar wind range of $\sigma = 10\text{-}100$ \cite{Beloborodov2020}. Given $\textbf{\emph{E}}_{\rm{peak}}\simeq\beta_{0}B_{0}\hat{\textit{y}}$ and $\textbf{\emph{B}}_{\rm{peak}}\simeq3B_{0}\hat{\textit{z}}$ at shock front when $\sigma\geq 1$ \cite{Plotnikov2018,Plotnikov2019} and combining with Eq. (2), the QED factor in the shock is 
$\chi=2\times10^{-16}\sqrt{\sigma\gamma_0^3 n_0}\simeq10^{-20}\gamma_0\nu$.

The precursor field is typically $\delta B \simeq (0.1-1) B_0$ \cite{Plotnikov2019}.  Thus, the normalized field strength of the precursor wave is estimated as 
$a_0= e\delta B/m_e\omega_{\rm{peak}}=(0.1-1)\gamma_0/3$.
Substituting $N$, $\chi$ and $a_0$ into Eq. (3), we get the criterion of  significant RR,
\begin{equation}
\gamma_0^2\nu\sim 3\times(10^{19}-10^{21})
\end{equation}

\noindent{From Eqs. (2) and (4), with typical FRB band $\nu = 10^{9–10}\;\rm{Hz}$, the requirements for effective RR  in shock emission are $\gamma_0 \sim 10^{4-6}$ and $\sigma n_0 = 1.37\times10^{9–11}/\gamma_0$. For highly magnetized shocks launched in magnetar winds, $\sigma\sim(10-100)$\cite{Beloborodov2020}, one obtains $n_0\sim10^{2-5}\ \rm{cm}^{-3}$.  Here $n_0$ and $\gamma_0$ are defined in the simulation frame (downstream rest frame). In the laboratory frame (upstream rest frame), the upstream density $n_{0,\text{u}}$ and shock Lorentz factor $\gamma_{\rm{s|u}}$ satisfy $n_{0,\text{u}}=n_0/\gamma_0$ and $\gamma_{\rm{s|u}}=2\gamma_0\sqrt{\sigma}$, yielding $n_{0,\text{u}}\sim10^{-4}-10\ \rm{cm}^{-3}$ and $\gamma_{\rm{s|u}}\sim10^{4-7}$. }

We employ a standard set of parameters as our starting point: $\gamma_0=1\times10^6$, $\sigma=10$, and $n_0=1\times10^3\;\rm{cm^{-3}}$, which correspond to $\nu=2.7\;\rm{GHz}$ and $\chi_0=2\times10^{-5}$. And the parameter scan across $\gamma_0=10^3-10^7$, $\sigma=1-20$, and $n_0=10^3-10^6\;\rm{cm^{-3}}$ are also explored in this work \footnote{As higher $\sigma$ values demand greater computational accuracy, duo to the constraints of computation power, the magnetization $\sigma$ in our simulations are limited to the range of $1-20$, which is narrower than the realistic range of $\sigma=10-100$ for magnetar winds.}. 


The simulation demonstrated that the RR effect manifests in two stages: counter-propagation with the precursor wave upstream, and coherent gyration at the shock front. In the former stage, electrons/positrons moving in the $-\hat{\textit{x}}$ direction collide with the precursor wave field, leading to collective oscillations. When the RR module is turned off, these oscillations cause only slight fluctuations in the momentum and energy, as shown in Fig. 1(a-c). However, after activating the RR effect, these oscillations induce significant radiation damping, leading to a rapid decay in the longitudinal momentum $\gamma\beta_{x}$ and energy until the RR effect cannot be triggered, as depicted in Fig. 1(d-f). 

\begin{figure}
\includegraphics[width=\columnwidth]{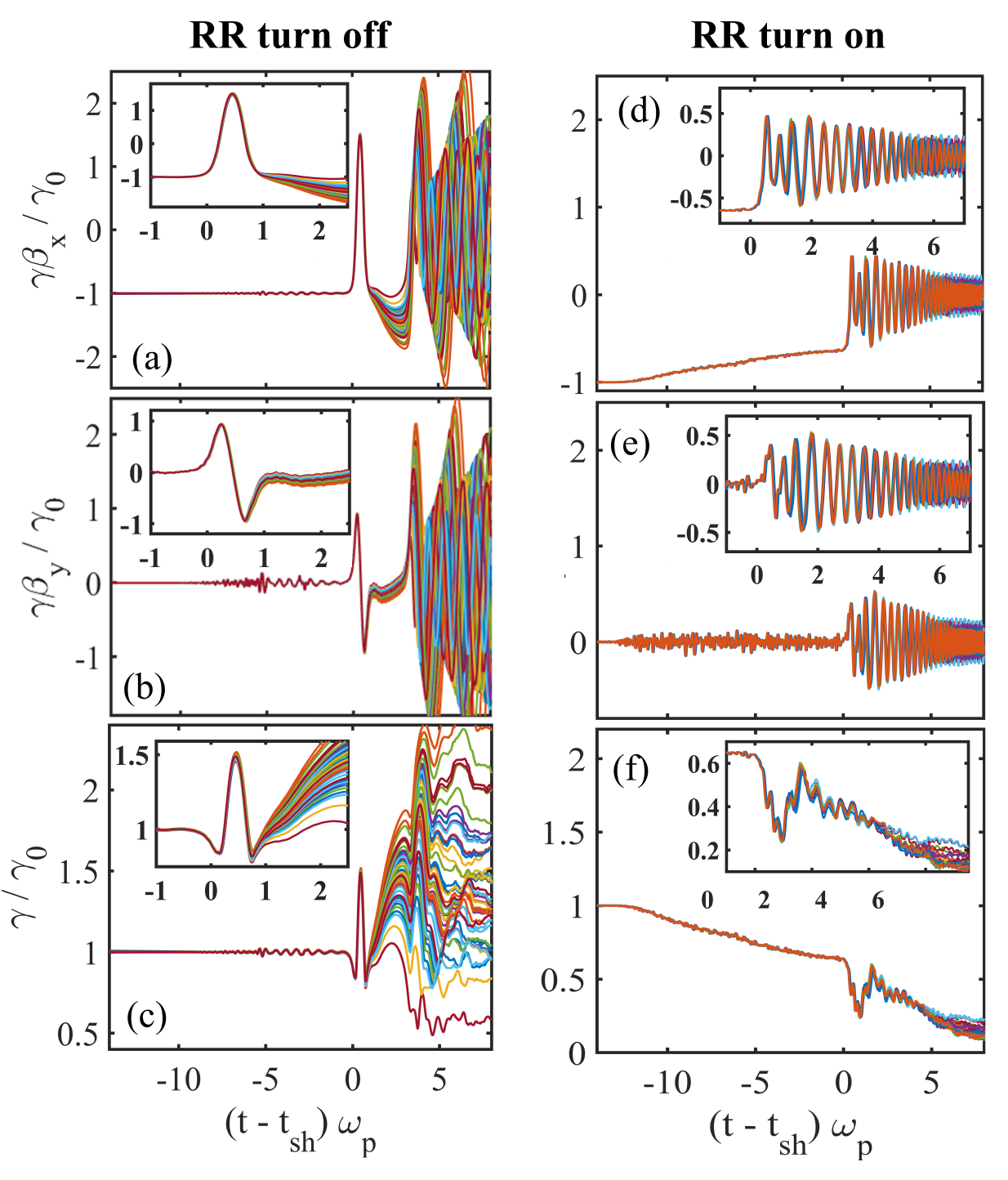}
\caption{\label{fig.1}
Evolution of the normalized longitudinal momentum $\gamma\beta_{x}$, transverse momentum $\gamma\beta_{y}$ and Lorentz factor $\gamma$ of the positron flow with RR turned off (a-c) and on (d-f). Here 100 adjacent positrons are chosen from the upstream and  $t_{\rm{sh}}$ is the moment when upstream particles entering the shock. The insets in (a-f) provide a magnified view of the coherent gyration cycles at the shock front. The initial parameter settings for the RR and no-RR cases are consistent: $\gamma_0=1\times10^6$, $\sigma=10$ and $n_0=1\times10^3\;\rm{cm}^{-3}$.
}
\end{figure}

In the latter stage, the injected flow encounters the shock and undergoes coherent gyration at the shock front. When the RR module is deactivated, the electron/positron flow thermalizes after completing one cycle of coherent gyration, and then transitions into the downstream region, as shown in Fig. 1(a-c). It is consistent with the $\gamma\beta_{x}-\gamma\beta_{y}$ distribution illustrated in Fig. 2(c). When the RR module is switched on, the energy damping induced by RR reduces the electron momentum significantly and shortens the electron cyclotron period $T \sim 2\pi\gamma m_{e}c/eB_{0}$. This allows the electrons to execute multiple coherent gyrations at the shock front prior to downstream entry, as shown in Fig. 1(d-f). During this process, the radial momentum continuously decreases under the radiation damping, producing a distinctive snail-shell pattern in the momentum trajectory, as illustrated in Fig. 2(d). It should be noted that the number of coherent gyrations at the shock front is not constant after the RR effect is triggered. However, it generally increases with the strength of the RR effect.

\begin{figure}
\includegraphics[width=\columnwidth]{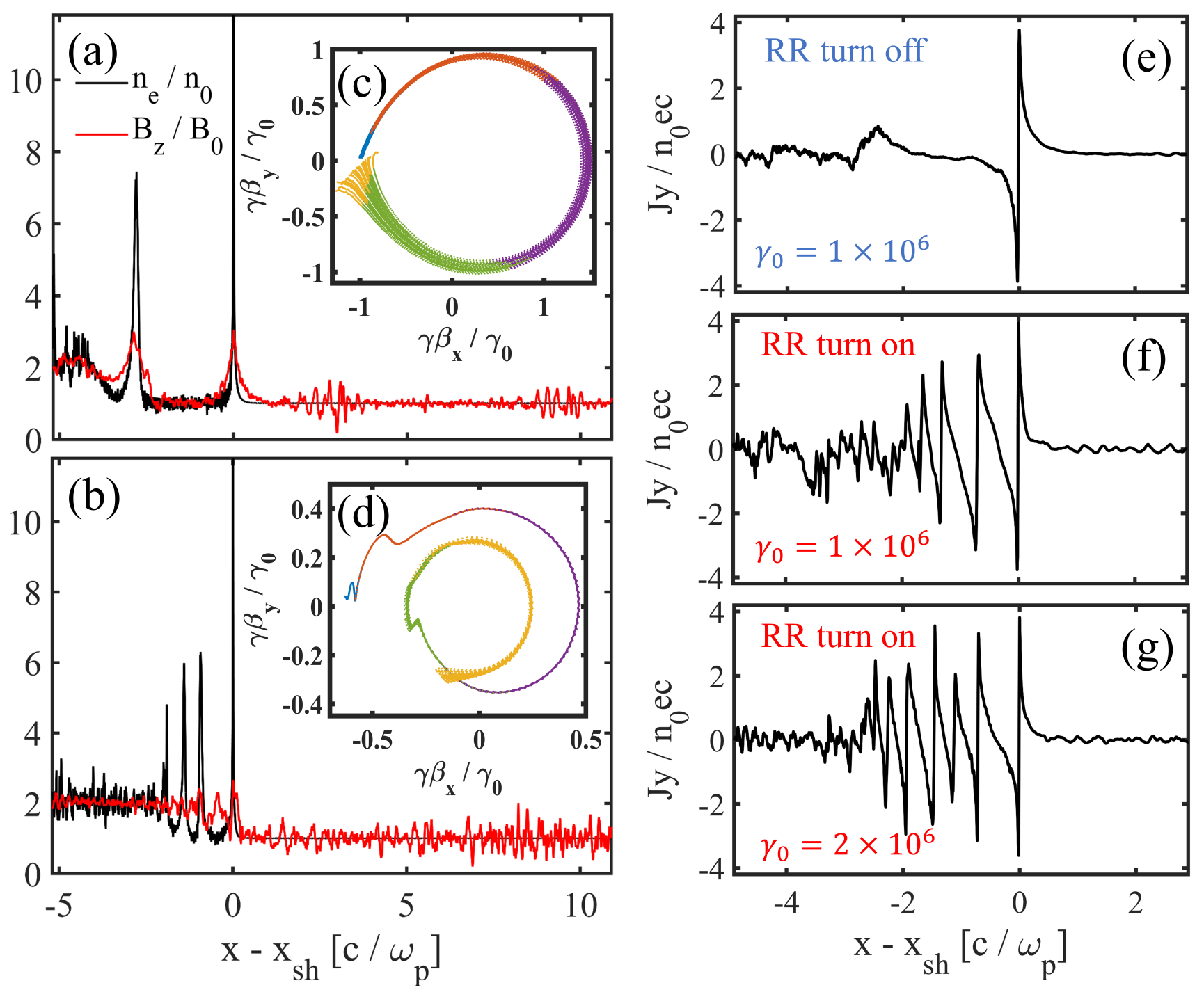}  
\caption{\label{fig.2}
Coherent gyration characteristics of shock front particle flow when RR is closed and open. (a) and (b) show the spatial distribution of density and magnetic field around the shock front with RR disabled and enabled, respectively. Black lines represent the normalized electron density $n_{e}/n_{0}$, and red lines represent the magnetic field $B_{z}/B_{0}$. Here the $\gamma_0=1\times10^6$, $\sigma=10$, $n_0=1\times10^3\;\rm{cm^{-3}}$ are consistent in both RR and no-RR case. (c) and (d) display sequential temporal snapshots of positron flow dynamics in momentum space during shock entry prior to thermalization, with blue, red, yellow, purple, and green markers representing progressively later time points at $20 \Delta t$ intervals. Distribution of the transversal current density $J_{y}$ with no-RR (e) and with RR where $\gamma_0=1\times10^{6}$ (f) and $\gamma_0=2\times10^{6}$ (g), where $\sigma=10$ and  $n_0=1\times10^{3}\;\rm{cm^{-3}}$ are fixed. 
}
\end{figure}

The RR effect-induced modifications substantially reshape the shock structure.  The coherent gyration at the shock front generates localized density peaks along the $x$-axis. Unlike the no-RR scenario, which typically exhibits only one or two density peaks at the shock front, the RR-mediated shock front shows multiple peaks resulting from the increased gyration cycles. The inter-peak density cavities are narrower, as compared between Fig. 2(a) and (b). As a result, the coherent radiation spectrum is drastically changed by the RR effect. The spectral distributions of $|\delta B_{z} |^2/B_{0}^{2}$ ahead of the shock are compared in Fig. 3(a) and (b). By correlating the changes with the $\gamma\beta_x-\gamma\beta_y$ trajectory of the particle cyclotron shown in Fig. 2(c) and (d), the key spectral features of RR-mediated shock radiation can be summarized as follows:

\begin{figure}
\includegraphics[width=\columnwidth]{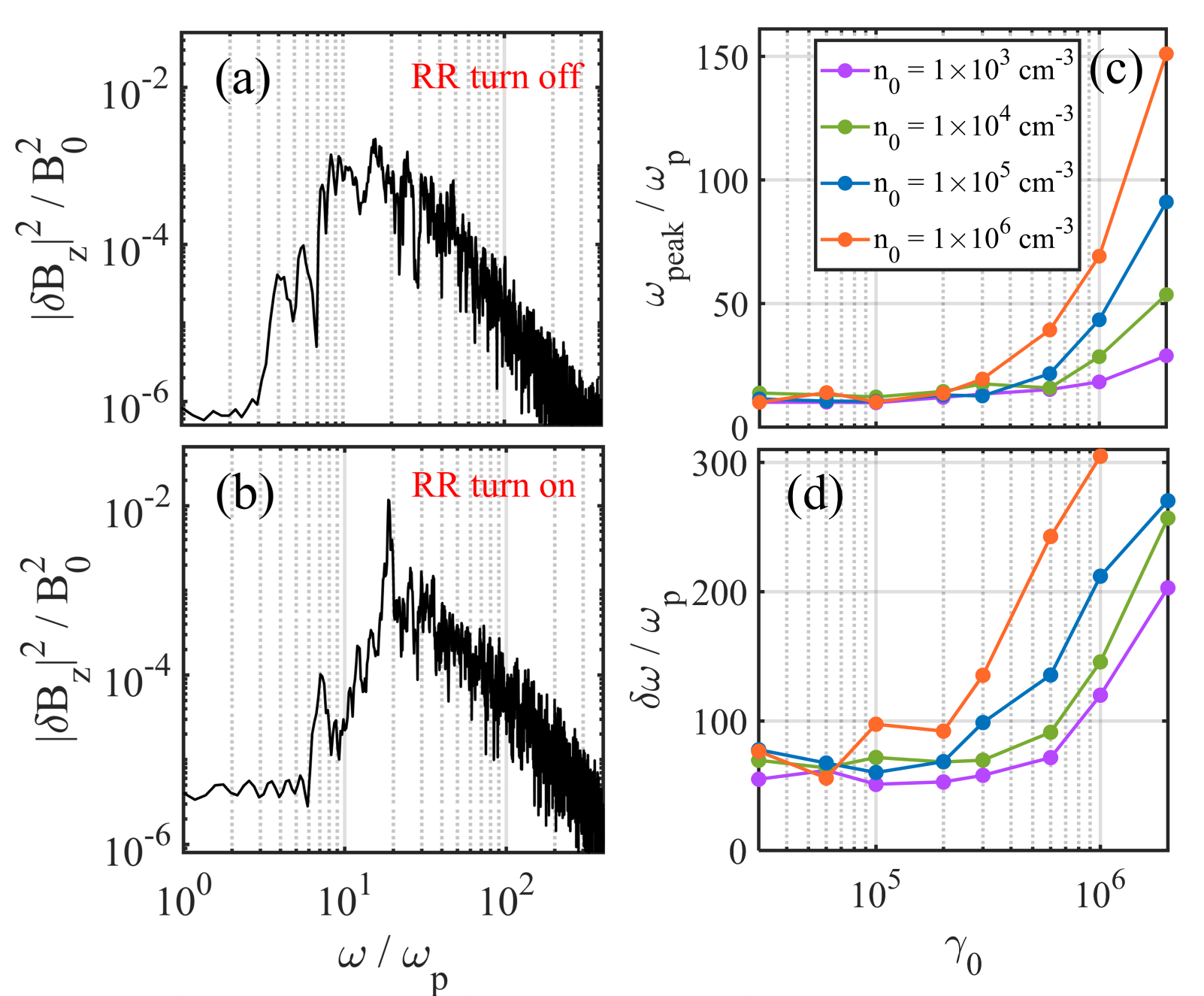}
\caption{\label{fig.3}  The spectrum of $|\delta B_{z} |^2/B_{0}^{2}$ ahead of the shock front with no-RR (a) and with RR (b), where $\gamma_0=2\times10^{6}$, $\sigma=10$ and  $n_0=1\times10^{2}\;\rm{cm^{-3}}$ are same in both (a) and (b). The spectra are normalized as $\int|\delta B_{z}(\omega)|^{2}/B_{0}^{2}d\omega=\langle\delta B_{z}^{2}\rangle/B_{0}^{2}$, $\delta B_{z}=B_{z}-B_{0}$. (c) and (d) represents the evolution of peak angular frequency $\omega_{\rm{peak}}$ and spectral width $\delta \omega$ in the $|\delta B_{z} |^2/B_{0}^{2}$ spectrum with $\gamma_0$ for $n_0=1\times10^3\;\rm{cm^{-3}}$ (purple), $1\times10^4\;\rm{cm^{-3}}$ (green), $1\times10^5\;\rm{cm^{-3}}$ (blue) and $1\times10^6\;\rm{cm^{-3}}$ (orange), where $\sigma=10$ is fixed.} 
\end{figure} 

(1) Upshifted peak frequency. As shown in Fig. 3(a), the spectrum in the absence of RR exhibits a cutoff frequency $\omega_{\rm{cutoff}} = 3.2\;\omega_p$ and a peak frequency $\omega_{\rm{peak}} = 14.5\;\omega_p$ at $\sigma=10$, consistent with the scaling relations $\omega_{\rm{cutoff}} \sim \sqrt{\sigma}\omega_p = \omega_c$ and $\omega_{\rm{peak}} \sim (3–4)\;\omega_c$ reported in previous  high-$\sigma$  studies \cite{Plotnikov2019,Sironi2021}. With RR included, as seen in Fig. 3(b), radiative cooling during precursor counter-propagation lowers $\gamma$ at the shock front, increasing $\omega_c = eB_0/\gamma m_e$. This up-shifts the peak frequency to $\omega_{\rm{peak}} = 18.65\;\omega_p$. Fig. 3(c) denotes the dependence of $\omega_{\rm{peak}}$ on $\gamma_0$ at fixed $\sigma$ and $n_0$. With increasing $\gamma_0$, RR effects become stronger, leading to a systematic increase in $\omega_{\rm{peak}}$. Similarly, raising the density $n_0$ also enhances the RR strength and further boosts the peak frequency.

(2) Broader frequency bandwidth. During the coherent cyclotron process, the particle flow continuously loses energy, leading to significant increase in radiation frequency and broadening of the spectrum. As shown in Fig. 3(d), the spectral bandwidth $\delta\omega$ grows with the $\gamma_0$ of the incident flow. Here, $\delta\omega$ is defined as the frequency range over which the intensity exceeds $I_\omega$, where $I_\omega$ satisfies $\log_{10}(I_{\omega})=[\log_{10}(I_{\rm{peak}})+\log_{10}(I_{\rm{cutoff}})]/2$.  $I_{\mathrm{peak}}$ and $I_{\mathrm{cutoff}}$ represent the intensities $|\delta B_z|^2/B_0^2$ of radiation spectrum at $\omega_{\mathrm{peak}}$ and $\omega_{\mathrm{cutoff}}$, respectively.

(3) Amplification of spectral peaks. The density cavity at the shock front, shown in Fig. 2(a), forms between two density peaks, corresponding to separated electron sheets. It acts as a mode selector for initial cyclotron radiation, analogous to a laser resonant cavity \cite{Plotnikov2019}. With RR effects, coherent gyrations of the electron flow generate additional density peaks, effectively forming multiple cavities as seen in Fig. 2(b). As RR induces a sufficient number of density peaks, the soliton structure at the shock front enhances mode selection and amplifies radiation near the peak frequency, leading to a sharp spectral peak in the radiation spectrum.

\begin{figure}
\includegraphics[width=\columnwidth]{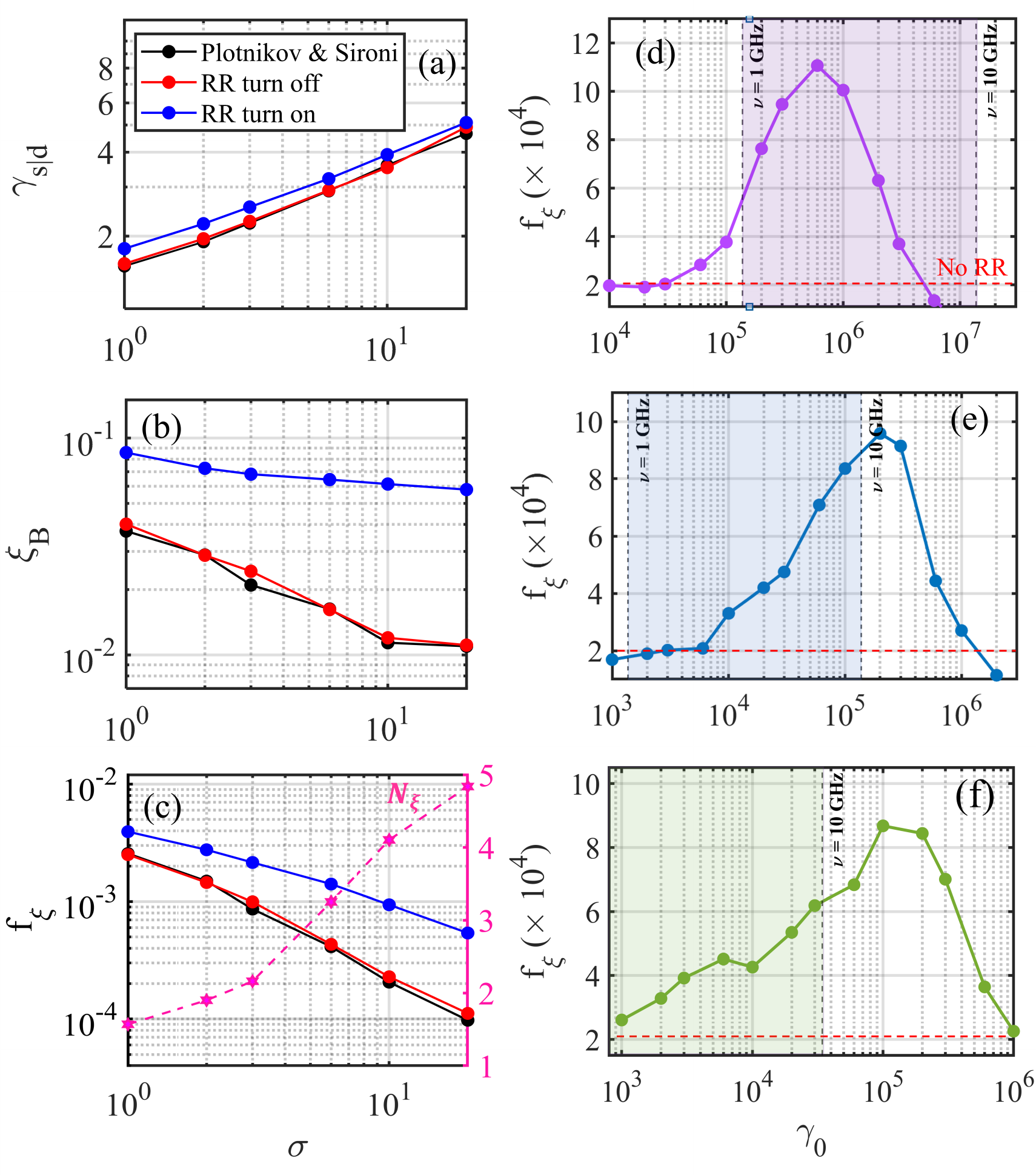}
\caption{\label{fig.4}
(a–c) display $\gamma_{\rm{s|d}}$, $\xi_B$, and $f_\xi$ versus $\sigma$ in the downstream frame for a fixed $\gamma_0 = 1 \times 10^6$ and $\sigma n_0 = 1 \times 10^4$, thus constant $\chi_0 \sim 2 \times 10^{-16} \sqrt{\gamma_0^3 \sigma n_0} = 2 \times 10^{-5}$. The black, red, and blue dots denote the results from previous maser shock simulation without RR [extract from the Fig. 3 in ref \cite{Plotnikov2019}], RR-deactivated shock and RR-activated shock, respectively. The magenta star in (c) represents the ratio of $f_\xi$ with RR turned on \textcolor{red}{to} that with RR turned off, i.e., $N_\xi = f_{\xi,\rm{RRon}} / f_{\xi,\rm{RRoff}}$. (d-f) shows the dependence of $f_{\xi}$ on $\gamma_{0}$ in RR-mediated shocks with $\sigma=10$, $n_{0}=1\times10^{3}\;\rm{cm^{-3}}$ (purple), $1\times10^{5}\;\rm{cm^{-3}}$ (blue) and $4\times10^{5}\;\rm{cm^{-3}}$ (green). The colored region shows the $\gamma_0$ range for coherent radiation peak frequencies falling within 1–10 GHz. The red dashed line marks the baseline of $f_\xi\sim2\times10^{-4}$ in the absence of RR.}
\end{figure}

Unlike the single $J_y$ current pulse generated by one gyration cycle shown in Fig. 2(e), multiple current pulses arise as the number of coherent cyclotron orbits rises. The increasing of $\gamma_0$ strengths the RR effect, progressively shortening the gyration period and thus yielding more coherent gyrations and additional current pulses, as illustrated in Figs. 2(f)–(g). Notably, even though RR depletes electron/positron energy, radiation from the multiple current pulses coherently superposes upstream, creating constructive interference that amplifies the coherent radiation.

Parameter scan with fixed QED factor $\chi$, the shock's lorentz factor measured in downstream rest frame $\gamma_{\rm{s|d}}$, radiation intensity $\xi_{B}$, and energy conversion efficiency $f_{\xi}$ with respect to $\sigma$ is illustrated in Fig. 4(a-c). Following previous shock emission studies \cite{Plotnikov2019,Sironi2021}, we define the $\xi_{B}=\langle\delta B_{z}^{2}\rangle/B_{0}^{2}$, and  $f_{\xi}=\xi_{B}\sigma(1+\beta_{\rm{s|d}})/[(1+\sigma)(\beta_{0}+\beta_{\rm{s|d}})]$ \footnote{The detailed explanation for the formula of $f_{\xi}$ can be found in ref\cite{Petroff2019}or the supplemental material}. Our results are consistent with those from previous shock radiation simulations \cite{Plotnikov2019} without RR. While similar trends are noticed in the RR case, obvious enhancement is observed with $\sigma$ ranging $1\sim20$. Particularly, both $\xi_{B}$ and $f_{\xi}$ show a 2-6 enhancement compared to normal shock conditions.

We also performed parameter scans over $\gamma_0 = 10^{3–7}$ and $n_0=10^{3-5}\ \text{cm}^{-3}$ with fixed $\sigma=10$ to evaluate the energy transferring efficiency $f_\xi$, as shown in Fig. 4(d–f). The influence of RR falls into three regimes: (1) RR-inactive regime. At low $\gamma_0$, radiative cooling is negligible and the shock evolution matches the non-RR case. The value $f_\xi \sim 2\times10^{-4}$ is nearly independent of $\gamma_0$, consistent with previous work at similar $\sigma$ \cite{Plotnikov2019,Sironi2021,Iwamoto2017,Gallant1992}. (2) RR-enhanced regime. With increasing $\gamma_0$, RR damping gradually boosts $f_\xi$ until it saturates at a factor of 4–6. This enhancement strengthens at higher magnetization, as shown in Fig. 4(c). For shocks within strongly magnetized magnetar winds with $\sigma=10-100$ \cite{Beloborodov2020,Sironi2021}, the enhancement is predicted to be even more pronounced. (3) RR-suppressed regime. At sufficiently high $\gamma_0$, extreme radiative cooling in the precursor region severely depletes particle energy, driving $f_\xi$ below the non-RR baseline. Meanwhile, increasing $n_0$ reduces the saturation value of $\gamma_0$ for RR-induced enhancement. As shown in Fig. 4(d–f), at fixed $\sigma=10$, with $\gamma_{0|\rm{u}}=2\sqrt{\sigma}\gamma_0\sim10^{4-7}$ and corresponding $n_{0,\rm{u}}=n_0/\gamma_0$ ranging from $10^{-3}$ to tens of $\rm{cm}^{-3}$, the RR-enhanced regime overlaps significantly with the $1–10$ GHz frequency interval. The upper part of the $n_{0,\text{u}}$ range overlaps considerably with the typical density range of magnetar winds\cite{Beloborodov2017,Beloborodov2020}, suggesting that the RR effect could be triggered in the highly relativistic portion of shock maser emission from magnetar winds associated with FRBs, and may also accompany high-energy shock evolution in more tenuous astrophysical environments.


Based on the above analysis, RR-modified coherent emission from strongly relativistic magnetized shocks in magnetar-wind environments produces a higher peak frequency, enhanced radiation intensity, and modified spectral morphology, including bandwidth broadening and pronounced spectral peaks. These features can be related with several FRB observations and may offer a new perspective on their coherent emission. For example, in the CHIME/FRB catalog \cite{Chen2022, Pleunis2021}, one‑off FRBs show statistically higher luminosities and broader bandwidths than repeating ones. Our scenario suggests that a brighter burst is more likely to trigger the RR effect, which in turn broadens its bandwidth. Some observed FRBs exhibit extremely narrow spectral structure near the peak frequency $\nu_{\rm{peak}}$, with $\Delta\nu/\nu_{\rm{peak}}\lesssim 0.1$ \cite{Kumar2021,Zhang2023,Zhou2022,Law2017,Yang2023}. In our model, the multi‑soliton‑cavity shock structure strongly amplifies the radiation at the peak frequency, producing a distinct peak. As shown in Fig. 3(b), the FWHM of the spectrum is $\delta \omega_{\text{FWHM}} \simeq 0.60\;\omega_p$, and $\omega_{\text{peak}} \simeq 18.65\;\omega_p$, corresponding to $\delta \omega_{\text{FWHM}} / \omega_{\text{peak}} = 0.032<0.1$. The rapid increase in coherent radiation efficiency after RR triggering  provides a useful comparison with the bimodal energy distribution exhibited in the repeating FRB 20121102A\cite{LiD2021}: the low-energy mode might arise from RR‑free shocks, while the high-energy mode could involve RR‑triggering shocks. Here the bursts in high‑energy mode also tend to have broader bandwidths than the low‑energy mode \cite{Zhong2022}. 

In conclusion, our study demonstrates the critical role of RR in reshaping coherent radiation from magnetized shocks. RR induces multiple coherent electron gyration cycles, leading to several-fold gains in radiation intensity and energy efficiency, along with a modified spectrum featuring an up-shifted peak, broaden bandwidth, and narrow spectral peak. These signatures can align with several statistical and spectral properties of FRBs, indicating that RR-modified shocks may be relevant in a subset of FRB shock-maser conditions. More broadly, our results establish a connection between relativistic shock physics, radiation reaction, and strong-field plasma processes, with implications for high-energy astrophysics and future laboratory probes of RR-related strong-field QED effects.

\hspace*{\fill}
\begin{acknowledgments}
We would like to thank the anonymous referees
for their helpful criticism and comments. We also thank Xue-Song Geng for supporting on numerical issues and Pablo J. Bilba for enlightening conversations. This work was supported by the National Science Foundation of China (Grant No. 12388102), the Strategic Priority Research Program of Chinese Academy of Sciences (No. XDB0890302), the CAS Project for Young Scientists in Basic Research (Grant No. YSBR060), the National Natural Science Foundation of China (No. 12473047), the National Key Research and Development Program of China (No. 2024YFA1611603), and the Yunnan Key Laboratory of Survey Science (No. 202449CE340002).
\end{acknowledgments}
\bibliography{FRB_RR}
\end{document}